\documentclass[12pt,preprint]{aastex}

\slugcomment{To Appear in The Astrophysical Journal}

\shorttitle{self-gravitating gaseous disks}
\shortauthors{M. A. Jalali and M. Abolghasemi}

\begin{document}

\title{SCALE-FREE EQUILIBRIA OF SELF-GRAVITATING GASEOUS DISKS
WITH FLAT ROTATION CURVES}

\author{Mir Abbas Jalali\altaffilmark{1}}
\affil{Institute for Advanced Studies in Basic Sciences,
P.O. Box 45195-159, Zanjan, IRAN \email{jalali@iasbs.ac.ir} }

\and

\author{Mehdi Abolghasemi}
\affil{Aerospace Research Institute, P.O. Box 15875-3885, Tehran, IRAN
\email{mehdi@ari.ac.ir}}

\altaffiltext{1}{Scientific Advisor, Aerospace Research Institute, Tehran, IRAN}

\begin{abstract}
We introduce exact analytical solutions of the steady-state hydrodynamic
equations of scale-free, self-gravitating gaseous disks with flat rotation
curves. We express the velocity field in terms of a stream function and
obtain a third-order ordinary differential equation (ODE) for the angular
part of the stream function. We present the closed-form solutions of the
obtained ODE and construct hydrodynamical counterparts of the power-law
and elliptic disks, for which self-consistent stellar dynamical models
are known. We show that the kinematics of the Large Magellanic Cloud
can well be explained by our findings for scale-free elliptic disks.
\end{abstract}


\section{INTRODUCTION}
Astrophysical disks are either stellar or gaseous. The existence of
non-axisymmetric equilibrium states of such systems has been attractive
both from theoretical and observational aspects. The search for
dynamical equilibria of stellar systems is known as the self-consistency
problem where one seeks for a positive phase space distribution
function (DF) that produces the density of the model and satisfies the
collisionless Boltzmann equation (Binney \& Tremaine 1987). The construction
of the DF can be done by analytical means or numerical exploration based on
Schwarzschild's orbit superposition method (Schwarzschild 1979; Kuijken 1993,
hereafter K93; Levine \& Sparke 1998; Teuben 1987; de Zeeuw, Hunter, \&
Schwarzschild 1987; Jalali \& de Zeeuw 2002, hereafter JZ). The main
difficulty linked to the self-consistency problem is the lack of the second
integral of motion for most planar models.

Gaseous disks, however, are investigated through solving the hydrodynamic
equations of compressible, self-gravitating fluids. There are several
works in the literature that deal with the stability of perturbed
axisymmetric gaseous disks (Heemskerk, Papaloizou, \& Savonije 1992;
Goodman \& Evans 1999; Lemos, Kalnajs, \& Lynden-Bell 1991).
The problem of non-axisymmetric systems is somewhat different
because our knowledge is limited even about equilibrium states. Following
a first-order linear theory, Syer \& Tremaine (1996, hereafter ST96)
constructed simple scale-free disks and approximately solved the
steady-state hydrodynamic equations for the components of the velocity
vector. They also proposed a nonlinear theory based on the calculation
of Fourier coefficients using Newton's method. A similar strategy was
adopted by Galli et al. (2001, hereafter G01) in the study of isothermal
molecular clouds and the fragmentation hypothesis of multiple stars.
Their nonlinear solutions, in contrast with ST96, were generated by
direct numerical integration of reduced ordinary differential
equations (ODE) and an iterative scheme to match the boundary conditions.
These methods of approach seem inefficient when one seeks for global
properties of highly non-axisymmetric equilibria and the transient response
of fluid disks in the vicinity of such states. Therefore, finding analytical
solutions for the hydrodynamic equations is of special interest even in
the steady-state conditions.

In this paper we explore exact non-axisymmetric solutions of self-similar
gaseous disks with flat rotation curves. In \S2 we express the governing
equations and reduce them to a third-order ODE in \S3 where we also introduce
our {\it closed-form} analytic solutions. We construct hydrodynamical
counterparts of the JZ models in \S4 and apply our results to the Large
Magellanic Cloud (LMC) in \S5. We show that the kinematics of the LMC can
be explained by our model elliptic disks. We end up with conclusions in \S6.

\section{STEADY-STATE HYDRODYNAMIC EQUATIONS}
We consider a thin disk of compressible, inviscid self-gravitating
fluid in a steady state (dynamical equilibrium). The hydrodynamic
equations can be written as the time-independent continuity equation
\begin{equation}
\label{eq::continuity}
\frac 1R {\partial \over \partial R}(R\Sigma v_R)+
\frac 1R {\partial \over \partial \phi}(\Sigma v_\phi)=0,
\end{equation}
and the momentum equations
\begin{eqnarray}
\label{eq::r-equation}
&{}& \!\! {1 \over R}{\partial \over \partial R}(R\Sigma v_R^2) +
{1 \over R}{\partial \over \partial \phi}(\Sigma v_R v_\phi)
- {\Sigma v_\phi ^2 \over R} = - {\partial p \over \partial R}
- \Sigma {\partial U \over \partial R}, \\
\label{eq::theta-equation}
&{}& \!\! \frac 1R {\partial \over \partial R}(R\Sigma v_R v_\phi) +
\frac 1R {\partial \over \partial \phi}(\Sigma v_\phi ^2) + {\Sigma
v_R v_\phi \over R} = - \frac 1R {\partial p \over \partial \phi}
- \Sigma \frac 1R {\partial U \over \partial \phi},
\end{eqnarray}
where $(R,\phi)$ are the polar coordinates,
${\bf v}=(v_R,v_\phi)$ is the velocity field in Eulerian
description, and $p$ is the pressure. The potential $U$ is
due to the self-gravity, which is related to the surface
density $\Sigma$ through the Poisson integral
\begin{equation}
\label{eq::phi-sigma}
U(R,\phi)=-G\int \!\! \int {\Sigma(R',\phi')R'{\rm d}R'
{\rm d}\phi' \over \sqrt{R^2+R'^2-2RR'\cos(\phi-\phi')}},
\end{equation}
where $G$ is the gravitational constant. In the following section,
we attempt to simplify the governing equations for scale-free disks
and solve the resulting equations by analytical means.

\section{NON-AXISYMMETRIC DISKS WITH LOGARITHMIC POTENTIALS}
\label{sec:logarithmic-disks}
Gaseous disks with logarithmic potentials possess the astrophysically
important property of a flat rotation curve. The circular gaseous disks
having this property are known as Mestel (1963) disks, the stability of
which was investigated by Goodman \& Evans (1999). Here we derive
{\it exact analytical} expressions for the physical quantities
of non-axisymmetric disks.

\subsection{Stream Function}
\label{subsec:stream-function}
We assume cuspy, scale-free potential--density pairs of the form
\begin{eqnarray}
U(R,\phi) \! &=& \! U_0 [2\ln R + g(\phi)], \nonumber \\
\Sigma (R,\phi) \! &=& \! {\Sigma_0 f(\phi) \over R},
\label{eq::potential-density-pairs}
\end{eqnarray}
where $U_0$ and $\Sigma _0$ are constants. The functions $f(\phi)$
and $g(\phi)$ are $2\pi$-periodic in $\phi$ because they are
continuous. We normalize the surface density so that
\begin{equation}
{1\over 2\pi}\int\nolimits_{-\pi}^{\pi}f(\phi){\rm d}\phi=1.
\end{equation}
This transfers the scaling of $\Sigma$ to $\Sigma _0$
(e.g., G01) and implies $U_0=\pi G \Sigma_0$. Given $f$,
Poisson's integral does not usually lead to a closed-form
expression for $g$ and one has to adopt a series solution.
Let us assume that $f(\phi)$ admits the Fourier expansion
\begin{equation}
f(\phi)=1+\sum_{\ell =1}^{\infty}
           \left ( C_{\ell}\cos \ell \phi +
                   S_{\ell}\sin \ell \phi \right ),
\end{equation}
where $C_{\ell}$ and $S_{\ell}$ are constant coefficients. Using
eqs (A1) and (A5) of ST96, we find the potentials corresponding
to single Fourier modes and superpose the results. This gives
\begin{equation}
g(\phi)=-2\sum_{\ell =1}^{\infty}
        \left ( {C_{\ell} \over \ell} \cos \ell \phi +
                {S_{\ell} \over \ell} \sin \ell \phi \right ).
\end{equation}

We prescribe $f$ and determine $g$ through the scheme mentioned above.
Then, we attempt to solve the momentum equations for $v_R$, $v_{\phi}$,
and the pressure $p$ in terms of $f$ and $g$.

The velocity components can be derived from the stream function
$\Psi$ as
\begin{equation}
\label{eq::stream-function-definition}
\Sigma v_R=-{1\over R}{\partial \Psi \over \partial \phi},~~
\Sigma v_\phi={\partial \Psi \over \partial R}.
\end{equation}
The scale-freeness of the model suggests the Ansatz
\begin{equation}
\label{eq::stream-fnc}
\Psi(R,\phi) \! = \Sigma _0 V_{\phi} \Big [\ln R + P(\phi) \Big],
\end{equation}
where $V_{\phi}$ is a constant reference velocity that will become the
circular velocity if the disk is axisymmetric ($f=1$, $P=0$). We utilize eqs
(\ref{eq::potential-density-pairs}) through (\ref{eq::stream-fnc}) and take
the curl of eqs (\ref{eq::r-equation}) and (\ref{eq::theta-equation}).
This eliminates the pressure $p$ and leaves us with the following ODE
for $P(\phi)$
\begin{equation}
\label{eq::p-equation-isothermal}
P^{\prime \prime \prime}-{2f^{\prime}\over f}P^{\prime \prime}
+\frac {1}{f^2} \big [2(f^{\prime})^2-ff^{\prime \prime}+f^2 \big ]
P^{\prime} = \beta f \big ( fg^{\prime}+2f^{\prime} \big ),
\end{equation}
where $\beta =U_0/V_{\phi}^2$ is a dimensionless parameter and the prime
denotes ${\rm d}/{\rm d}\phi$. The solution $P(\phi)$ will result in the velocity
field of our planar self-gravitating fluids. Physical solutions of
(\ref{eq::p-equation-isothermal}) correspond to the condition
${\bf v}(R,\phi) = {\bf v}(R,\phi +2\pi)$, which ensures the continuity of
velocity field. Hence, we seek for $2\pi$-periodic solutions of $P(\phi)$.

\subsection{Velocity Components}
\label{subsec:velocity-component}

Carrying out a change of dependent variable as
(this is indeed Liouville's transformation explained
in Appendix A)
\begin{equation}
\label{eq::liouville-transform}
P' = w(\phi){f(\phi)\over f(0)},
\end{equation}
eq. (\ref{eq::p-equation-isothermal}) reads
\begin{equation}
\label{eq::w-equation}
w^{\prime \prime}+w={\beta f(0)} F_0(\phi),~~
F_0=fg'+2f'.
\end{equation}
The particular solution of (\ref{eq::w-equation})
is then obtained as
\begin{equation}
w(\phi)\!=\! {\beta f(0)} \Big [ a_0
\sin \phi \! + \! b_0 \cos \phi \! + \!
\int\nolimits_{0}^{\phi}F_0(s) \sin (\phi -s){\rm d}s \Big ],
\label{eq::w-solution}
\end{equation}
where the (dimensionless) constant coefficients $a_0$
and $b_0$ are given by
\begin{equation}
\left \{ \!\!\!
                 \begin{array}{c} a_0 \\ b_0
                 \end{array}
       \!\!\! \right \} \!=\! -{1\over \pi}\!
                              \int\nolimits_{0}^{2\pi}
\!\! \left \{ \!\!\!
                 \begin{array}{c} \sin \phi \\ \cos \phi
                 \end{array}
       \!\!\! \right \}  {\rm d}\phi \!\!
\int\nolimits_{0}^{\phi} \!\! F_0(s)\sin (\phi -s){\rm d}s.
\end{equation}
The function $w(\phi)$, and therefore ${\bf v}$, will be
$2\pi$-periodic if we impose the boundary conditions
$w(\phi+2\pi)=w(\phi)$ and $w^{\prime}(\phi+2\pi)=w^{\prime}(\phi)$
that lead to the necessary conditions
\begin{equation}
\label{eq::necessary-conditions-w}
\int\nolimits_{0}^{2\pi} \!\! F_0(\phi)\sin \phi {\rm d}\phi =
\int\nolimits_{0}^{2\pi} \!\! F_0(\phi)\cos \phi {\rm d}\phi = 0.
\end{equation}
Using (\ref{eq::liouville-transform}) and
(\ref{eq::stream-function-definition}), the velocity
components are readily determined as
\begin{eqnarray}
\label{eq::velocity-field-isothermal}
v_R \!\!\!\! &=& \!\!\!\! - \beta V_{\phi} \Big [ a_0 \sin\phi \!
+ \! b_0 \cos \phi \! + \! \int\nolimits_{0}^{\phi}
F_0(s)\sin (\phi -s){\rm d}s \Big ],   \nonumber \\
v_{\phi} \!\!\!\! &=& \!\!\!\! {V_{\phi} \over f(\phi)},
\end{eqnarray}
from which one can obtain $P(\phi)$. It is now straightforward
to determine the pressure $p$ from either (\ref{eq::r-equation}) or
(\ref{eq::theta-equation}) as
\begin{equation}
\label{eq::isothermal-pressure}
p={\Sigma _0 V_{\phi}^2 \over R f(\phi)} \left [\beta A(\phi)-1 \right ]=
c_s^2 \Sigma,
\end{equation}
where
\begin{equation}
\label{eq::sound-speed}
c_s^2={V_{\phi}^2 \over f(\phi)^2}[\beta A(\phi)-1],
\end{equation}
is the square of the sound speed (at the azimuth $\phi$) and
\begin{equation}
A(\phi) = f(\phi) \Big [ (b_0 \sin \phi - a_0 \cos \phi) +
2f(\phi)-\int\nolimits_{0}^{\phi} F_0(s)\cos (\phi -s){\rm d}s
\Big ].
\end{equation}
Sound waves are stable if $c_s^2>0$. i.e., $p>0$. Define
$A_{\rm min}=\min \left \{ A(\phi): 0\le \phi \le 2\pi \right \}$.
The necessary and sufficient condition for the positiveness of
$p$ will be
\begin{equation}
0<{1 \over \beta }<A_{\rm min}.
\end{equation}
Apart from constraining the admissible values of $\beta$,
this condition may put some restrictions on the model parameters.
As it can be seen in (\ref{eq::velocity-field-isothermal}),
$v_{\phi}$ is maximum (minimum) where the surface density is
minimum (maximum). This result has previously been used in the
construction of DFs of stellar systems using Schwarzschild's
orbit superposition method (e.g., K93; JZ).

Our scale-free analytical solutions are general. They are independent
of special angular variation of surface density and its corresponding
potential. Our solutions can be used even in the presence of an external
force that falls off as steep as $R^{-1}$. Such a force can be that of a
magnetic field or the gravitational influence of a dark matter halo.
These effects can enter in our equations through $F_0(\phi)$.

The most important point related to our solutions is the existence
of transonic flows. According to (\ref{eq::sound-speed}), it is
possible to obtain entirely subsonic, transonic, and entirely
supersonic flows depending on the choice of $\eta
=1/\sqrt{\beta A_{\rm min}}$ in the range $0<\eta <1$. Let us
define the Mach number
\begin{equation}
\label{eq::Mach-number}
{\cal M}=\frac 1{c_s}\sqrt {v_R^2+v_{\phi}^2},
\end{equation}
which is a function of the azimuthal angle $\phi$ in our disks. For
convenience, we use the variable
\begin{equation}
\label{eq::M0}
{\cal M}_0(\phi)={{\cal M}-1\over {\cal M}+1},
\end{equation}
for the classification of flows. Disks with ${\cal M}_0(\phi)>0$
(${\cal M}_0(\phi)<0$) for $\forall \phi \in [0,2\pi]$ are
entirely supersonic (subsonic) and our steady-state models are
credible. Flows for which ${\cal M}_0(\phi)$ switches sign along a
streamline, are transonic. In such a circumstance, steady-state
models become problematic as explained below. In principle,
transition from subsonic to supersonic speeds (along a streamline)
can take place smoothly, but the reverse is impossible without
assuming a shock wave. Because fluid elements in the subsonic part
of a streamline, have not information from the arrival of the
particles in the supersonic upstream. Therefore, they suddenly
find themselves collided with these particles. After a while,
these interactions develop a shock wave and the speed of
supersonic stream discontinuously decreases to a subsonic value.
This process is inherently time-dependent and it increases entropy
and generates heat. Thus, we must cautiously deal with any transonic
solution. Galli et al. (2001) also argue that models with transonic
flows will most likely be unstable. Despite these arguments, there
is an exceptional case in nature where a supersonic flowfield is
isentropically (and of course smoothly) compressed from supersonic
to subsonic speeds. That is the famous {\it conical flow} detailed
description of which can be found in Anderson (1990, chapter 10).
In this very special case, again, a shock wave still presents but
the {\it sonic line} does not coincide with the shock ray. The
latter completely lies in the supersonic region.

In any case, it seems reasonable to exclude transonic solutions from
our results. A stability analysis will surely provide much insight into the
problem. The instabilities of our non-axisymmetric disks will be resolved
in a future paper.

For axisymmetric disks, the parameter $\eta$ is obtained in terms of
Mach number through
\begin{equation}
\eta = {{\cal M}\over \sqrt{1+{\cal M}^2}}.
\end{equation}
In this case, ${\cal M}$ is independent of $\phi$ and the
transition from subsonic to supersonic flows occurs at
$\eta=1/\sqrt{2}$.

\section{EXAMPLES}
\label{sec:examples}
We now apply our analytical tool to certain non-axisymmetric
models for which we have either observational evidence or
stellar dynamical counterparts.

\subsection{Simple Disks}
\label{sec:simple-disks}
The simplest form of non-axisymmetric disks is obtained
when a circular disk is perturbed by a single Fourier mode as
\begin{equation}
f(\phi)=1 + f_m \cos m\phi,
\end{equation}
where $0 \le \vert f_m \vert <1$.
The corresponding $\phi$-dependence of the potential will become
\begin{equation}
g(\phi)=-\frac 2m f_m \cos m\phi.
\end{equation}
One can readily verify that the condition (\ref{eq::necessary-conditions-w})
is satisfied for all $m \ge 1$ and from (\ref{eq::velocity-field-isothermal})
we find
\begin{equation}
v_R=\beta V_{\phi} f_m \Big ( {f_m \over 4m^2-1}
\sin 2m\phi - {2 \over m+1}\sin m\phi \Big ).
\end{equation}
It is not difficult to show that $A(\phi)$ is positive for our
simple disks when $0 \le \vert f_m \vert <1$. These disks are
analogous to the multiple-lobed disks of G01. But they are not the
same. The difference can be traced back to the logic of solving
the hydrodynamic equations. In fact, hydrodynamic equations for
self-gravitating disks are four equations for the five unknowns
$\Sigma$, $U$, $v_R$, $v_{\phi}$ and $p$. Thus, we are free to
specify one of these variables, or put an extra constraint on
them. Galli et al. (2001) follow the latter approach and assume an
equation of state in the form $p=a^2\Sigma$ where $a$ is the
constant sound speed. They numerically solve their problem for
$\Sigma$, $v_{\phi}$, and $v_R$, and obtain $U$ through Poisson's
equation. Our disks, however, are constructed relying on another
approach. We prescribe the surface density $\Sigma$ and find $U$
using Poisson's integral. Then, we determine $v_R$, $v_{\phi}$,
and $p$. Our disks, in contrast to G01, are not isothermal [because
the sound speed is not constant according to (\ref{eq::sound-speed})].
Nonetheless, our solutions are exact and we never face the convergence
problem as G01.

In Figure 1, we have surveyed the $(\eta,f_m)$ parameter space of
our simple disks for $\pi G=1$, $\Sigma _0=1$, and several choices
of the wavenumber $m$. For each $m$, we obtain a curve (showed by
solid, dashed, and dash-dotted lines) which is the boundary of models
with transonic flowfields. The region to the up side of each curve
(indicated by the letter B) corresponds to the models that are
transonic along a typical streamline. Below the boundary curve, we
obtain entirely subsonic and entirely supersonic regions.
Excluding transonic solutions (they are unphysical due to
arguments in \S\ref{subsec:velocity-component}), highly
non-axisymmetric disks can be constructed in the subsonic region
as the wavenumber $m$ is increased. The maximum elongation of the
lopsided mode ($m=1$) occurs for $f_m \approx 0.5$ and $\eta
\approx 1/\sqrt{2}$. Whatever the value of $m$ may be, the maximum
value of $\eta$ is $1/\sqrt{2}$ for entirely subsonic models.
Entirely supersonic models are approximately circular with $f_m \la 0.2$.

\subsection{Bisymmetric Disks}
\label{sec:bisymmetric}
There are two interesting families of bisymmetric disks for which
self-consistent stellar dynamical models are known. They are the
power-law and elliptic disks (JZ). Disks whose equipotentials are
similar concentric ellipses, are called the power-law disks. They are
planar analogs of the logarithmic ellipsoidal models (e.g., Binney 1981;
Evans, Carollo \& de Zeeuw 2000). The power-law disks with flat rotation
curves take the potential of the form (\ref{eq::potential-density-pairs})
with
\begin{equation}
\label{eq::potential-power-law-disk}
g(\phi)=\ln (q^2\cos ^2\phi +\sin ^2\phi),
\end{equation}
where $q$ is the axial ratio of equipotentials. The angular
dependence of the surface density corresponding to
(\ref{eq::potential-power-law-disk}) has been derived in JZ as the
following convergent series
\begin{eqnarray}
\label{eq::density-power-law-disk}
f(\phi) \!\! &=& \!\! 1+\sum_{{\rm even}~\ell>0}^{\infty}
f_{\ell}\cos \ell \phi                \\
\label{eq::coefficient-power-law}
f_{\ell} \!\! &=& \!\! \sum_{n=1}^{\infty} \left \{
{(1-q) ^n \delta _{k\ell} \over 2^{n-1}}+\sum_{m=0}^{n-1}
{(1-q) ^n \beta ^{(n)}_m \over 2n} \Big [
i\delta _{i\ell}+j\delta _{j\ell} \Big ] \right \},   \nonumber
\end{eqnarray}
with $k=2n$, $i=n+m$, and $j=n-m$. In this relation, $\delta _{rs}$ is the Kronecker delta
and we have defined $\beta^{(n)}_m=0$ if $n-m$ is odd, and
\begin{eqnarray}
\beta^{(n)}_m \!\! &=& \!\! \left\{
\begin{array}{ll} \frac 1{2^{n-1}} \! \left ( \!\!\!\!
                          \begin{array}{c}  n \\
                                        {\displaystyle \frac{n-m}2}
                          \end{array} \!\!\!\! \right ),  &m \not=0, \\
                  \null & \null \\
                  \frac 1{2^n} \! \left ( \!\!\!\!
                          \begin{array}{c} n \\
                                       {\displaystyle     n/2}
                          \end{array} \!\!\!\! \right ), & m=0,  \\
\end{array} \right.
\label{eq:beta-potential}
\end{eqnarray}
if $n-m$ is even.

When the isodensity contours are similar concentric ellipses,
we attain elliptic disks with
\begin{equation}
\label{eq::density-elliptic-disk}
f(\phi)=(q^2\cos ^2\phi+\sin ^2\phi)^{-1/2},
\end{equation}
where $q$ is the axis ratio of isodensity contours. Stellar elliptic disks
were constructed (for the first time) in K93 and then were examined for
self-consistency by the curvature condition in JZ. The potential function
associated with (\ref{eq::density-elliptic-disk}) was determined by
Evans \& de Zeeuw (1992, eq. 5.2) in terms of an indefinite integral,
but a practical way is to generate it using the scheme presented in
\S\ref{subsec:stream-function}. We find
\begin{eqnarray}
\label{eq::potential-elliptic-disk}
U(R,\phi) \!\! &=& \!\! 2\pi G \Sigma _0 \Big ( \ln R-
\sum_{{\rm even}~\ell>0}^{\infty}{f_{\ell}\over \ell}\cos \ell \phi \Big ),  \\
\label{eq::coefficient-elliptic}
f_{\ell} \!\! &=& \!\! \sum_{n=1}^{\infty}\sum_{k=0}^{n}\sum_{m=0}^{n}
{(1-q) ^n \Gamma (\frac 12 +k)\Gamma (\frac 12 +n-k)
\beta ^{(n)}_m (\delta _{i\ell}+\delta _{j\ell})
\over 2\pi k! (n-k)!},        \nonumber
\end{eqnarray}
with $i=2k-n+m$ and $j=2k-n-m$. All the series introduced so far, rapidly
converge using only the first few terms.

For instance, we have computed the isodensity contours and
streamlines of our bisymmetric disks for $q=0.5$, $\pi G=1$, and
$\Sigma _0=1$, which result in $A_{\rm min}=0.854$ and $A_{\rm
min}=3.041$ for the power-law and elliptic disks, respectively.
Figure 2 shows the results of our calculations for several choices
of $V_{\phi}$. Solid lines are the isodensity contours and dashed
lines are streamlines. For the small values of $V_{\phi}$, the
pressure increases and dominates centrifugal forces. Therefore,
streamlines become highly dimpled in both disks (Panels 2{\em a}
and 2{\em c}). As the value of $V_{\phi}$ is increased
($V_{\phi}^2 \rightarrow U_0 A_{\rm min}$, i.e., $\beta
\rightarrow 1/A_{\rm min}$), centrifugal forces overwhelm the
pressure forces and keep the fluid particles away from the center.
This causes streamlines to become rounder (Panels 2{\em b} and
2{\em d}). Numerical computations show that $A(\phi)$ ($0\le \phi
\le 2\pi$) is positive for the power-law and elliptic disks in the
range $0\le q \le 1$. Therefore, it is always possible to find a
physical value for $V_{\phi}$. Figure 3 shows the $(\eta,q)$
parameter space of the power-law and elliptic disks. The region
filled by dots corresponds to models with transonic flowfields.
They are most likely unphysical models as we discussed in
\S\ref{subsec:velocity-component}. Hence, we exclude them from our
solution space. As Figure 3 shows, supersonic elliptic disks can
be more elongated than supersonic power-law disks. In both cases
the maximum value of $\eta$ for subsonic flows is equal to
$1/\sqrt{2}$. This is a generic property of our self-gravitating
gaseous disks inherited from simple disks.

Another remarkable point is the limit of the axis ratio $q$ for
plausible models (transonic models excluded). The lower bound of
$q$ is almost 0.4 in elliptic disks but in the case of power-law
disks one can see a narrow subsonic region with $q<0.08$. In the
limit of $q\rightarrow 0$, these highly elongated models converge
to a needle. The results for elliptic disks are similar to the
findings of JZ for stellar disks. In fact, stellar elliptic disks
with $q\la 0.5$ are non-self-consistent. On the evidence of Figure
3{\em a}, gaseous power-law disks exhibit significantly different
features compared to their stellar counterparts that cannot be
self-consistent if $q<0.707$ (JZ).

Our calculations show that the gas pressure of elliptic disks
approximately takes an axisymmetric distribution when $V_{\phi}$
is small. As a result, the line-of-sight velocity dispersion
$\sigma=\langle v_{\rm los}^2\rangle -\langle v_{\rm los}\rangle^2$
will inversely be proportional to $f(\phi)$. This property does not
depend on the choice of $q$ and it predicts a correlation between the
photometric and spectroscopic data of elliptic disks when centrifugal
forces are dominated by internal stresses.

\section{IMPLICATIONS FOR THE KINEMATICS OF THE LMC}
In his recent study of the LMC, van der Marel (2001) has shown that
the LMC disk is elongated and its isophotal lines are well fitted by
ellipses. Since the LMC has an approximately flat rotation curve, our
elliptic disks (presented in \S\ref{sec:bisymmetric}) can be applied to
model the LMC disk.

The most recent studies of the kinematics of gas and discrete tracers
(like carbon stars) of the LMC (Kim et al. 1998; Alves \& Nelson 2000)
show that the line of maximum velocity gradient is in the range
$143^\circ<\Theta _{\rm max}<183^\circ$, which differs from the angular
position of the line of nodes ($\Theta=122^{\circ}.5$) by
$20^{\circ}<\Theta_{\rm max}-\Theta<60^{\circ}$. Noting the fact that
the intrinsic ellipticity of the LMC disk is $\epsilon = 0.312$ in
outer regions, van der Marel (2001) argues that
$\Delta \Theta=\Theta _{\rm max}-\Theta \approx 20^{\circ}$ is possible but
it is hard to believe that $\Delta \Theta$ can be as large as $60^{\circ}$.
He ascribes the excess of computed misalignment to the center-of-mass and
rigid-body motions of the LMC. A more accurate kinematic analysis of the
LMC predicts a value of $\Delta \Theta$ as small as $7.4^{\circ}$
(van der Marel 2002, private communication). To provide a theoretical
framework for these observations, we have modeled the outer region of
the LMC using our elliptic disks.

We have calculated the position angle of the maximum line-of-sight
velocity of our elliptic disks, given the viewing angles
$\Theta=122^{\circ}.5$ and $i=34^{\circ}.7$ that were determined
by van der Marel \& Cioni (2001). Our results have been shown in
Figure 2. This figure illustrates the variation of $\Delta \Theta$
versus $\eta$ for several choices of $\epsilon=1-q$ including
$\epsilon=0.312$, which is the intrinsic ellipticity of the LMC at
large radii. A discontinuity occurs in the graph of $\Delta
\Theta$ (let us say at $\eta=\eta_ {\rm cr}$) because of a sudden
change in the morphology of streamlines. For $\eta<\eta_{\rm cr}$,
streamlines are dimpled. The graph for $\epsilon=0.312$ (filled
squares) shows that $\Delta \Theta$ varies between $0^{\circ}$ and
$30^{\circ}$ when streamlines are not dimpled. Interestingly, this
result is in harmony with the predictions of van der Marel (2001).
It is remarked that for $\epsilon =0.312$ we obtain $\eta _{\rm
cr}=0.36$. As $\eta \rightarrow 1$, streamlines become almost
circular independent of the value of $\epsilon$. In such a
circumstance, the position angle of the maximum line-of-sight
velocity coincides with that of the line of nodes, i.e. $\Delta
\Theta \rightarrow 0$, as expected. Based on the available
observational data and our results for $\epsilon=0.312$, we
conclude that our scale-free elliptic disks (with flat rotation
curves) can explain the kinematics of the outer region of the LMC
if $\eta$ is chosen in the range $0.36<\eta <0.72$. The upper
bound on $\eta$ comes from the fact that our transonic solutions
are not reliable answers of steady-state equations. For
$\Delta \Theta =7.4^{\circ}$, we find $\eta \approx 0.54$,
which is well within the acceptable range. Moreover,
models with $\eta \rightarrow 1$ (they are supersonic) lead to
very high fluctuations of the pressure, and therefore, velocity
dispersion. This does not seem to be relevant to a stable model.
We also rule out the possibility of $\eta <\eta _{\rm cr}$, for it
gives $\Delta \Theta <0$, which is not observed. Furthermore, the
condition $p\rightarrow 0$ at the boundary of the LMC disk is
inconsistent with a pressure-dominated model.

\section{CONCLUSIONS}
Exact equilibrium solutions of self-gravitating systems are rare.
Two classes of such solutions were found by Medvedev \& Narayan (2000)
for three-dimensional, axisymmetric isothermal systems.
A generalization of these solutions for non-axisymmetric
rotating systems was then introduced by Shadmehri \& Ghanbari (2001).
Although the density distribution of their models is non-axisymmetric,
the velocity field has only the toroidal component. This makes their
models too simple to be used in the study of fully three-dimensional
non-axisymmetric distributions of matter.

In a systematic search for non-axisymmetric equilibrium solutions,
we started our study from gaseous disks because they inherit many properties
of three-dimensional systems. Using the benefit of the existence of a
stream function, we were successful in reducing the governing equations
of scale-free models to an incomplete linear ODE for the angular part of
the stream function. It then became clear that Liouville's transformation
can reduce the obtained ODE to a second-order inhomogeneous ODE, which is
similar to that of a forced harmonic oscillator. This was striking because
it led us to derive closed-form solutions for the radial velocity component,
stream function and the pressure $p$. In fluid mechanics, transport equations
are written for mass, momentum and energy. While the pressure occurs in
governing equations, there is no extra independent equation in order
to determine $p$. The problem is rather difficult in the presence of
an anisotropic stress tensor, which is encountered in turbulence
and Jeans' equations of stellar hydrodynamics. Taking the curl of
momentum equations is a trick that temporarily removes the pressure
from our calculations of the stream function. But, one should take
care of the sign of $p$ once it is eventually extracted from momentum
equations. Only a positive pressure distribution is physical because
a fluid medium can resist against a compressive stress not a tensional
one.

Non-axisymmetric stellar disks have necessarily anisotropic DFs and
pressure tensors. This point is the main difference between the physics
of a steady-state gaseous disk and its stellar counterpart described
by Jeans' equations. Nevertheless, isotropic pressure distribution is
not an obstacle for the existence of non-axisymmetric,
self-gravitating gaseous disks as we showed in \S\ref{sec:examples}.

Our scale-free elliptic disks were used to model the outer part of
the LMC disk. The main restriction of our modeling is the isotropy
of the stress tensor. We speculate that our results will not be
changed in a significant way if the model is stellar with an
anisotropic pressure. The other important issue is the
perturbation of the Galaxy. The tidal field of the Galaxy induces
a force whose $R$-distribution is totally different from that of a
logarithmic potential field. However, if the tidal field of the
Galaxy was dominant, the rotation curve of the LMC would show
substantial deviation from flatness. But, we know that the
rotation curve of the LMC is almost flat. Therefore, our isolated
models provide acceptable {\it zeroth-order} solutions. Such a
modeling of the LMC is a quick route for interpreting the
observational data. It is obvious that taking the tidal effects
into account, will destroy the enjoyable scale-free nature of our
disks.

\acknowledgments We thank Tim de Zeeuw for useful comments and
drawing our attention to the possible application of our findings
to the LMC. We are indebted to an anonymous referee for
enlightening suggestions that helped us to substantially improve
the presentation of our results.

\appendix
\section{LIOUVILLE'S TRANSFORMATION}
Consider the following non-autonomous, linear, ordinary differential
equation for $Z$ in terms of $t$:
\begin{equation}
\label{eq::linear-ODE}
{{\rm d}^2Z \over {\rm d}t^2}+Q_1(t){{\rm d}Z \over {\rm d}t}+
Q_2(t) Z=Q_0(t).
\end{equation}
Using the transformation (Bellman 1966, page 82; Verhulst 1996,
problem 8-2)
\begin{equation}
\label{eq::Liouville}
Z(t)=w(t) \exp \left (-\frac 12 \int\nolimits_{0}^{t} \!\!
Q_1(\tau){\rm d}\tau \right ),
\end{equation}
which has been introduced by Liouville, eq. (\ref{eq::linear-ODE})
becomes
\begin{equation}
{{\rm d}^2 w\over {\rm d}t^2} + \Big ( Q_2 - \frac 12
{ {\rm d}Q_1 \over {\rm d}t } - \frac 14 Q_1^2 \Big )w =
Q_0(t)\exp \left (\frac 12 \int\nolimits_{0}^{t} \!\!
Q_1(\tau){\rm d}\tau \right ). \label{eq::reduced-linear-ODE}
\end{equation}
The study of this equation is easier than (\ref{eq::linear-ODE}).

We assume $P^{\prime}$ as the dependent variable. This puts
eq. (\ref{eq::p-equation-isothermal}) in the form of
(\ref{eq::linear-ODE}). We then apply eq. (\ref{eq::Liouville})
and obtain eqs (\ref{eq::liouville-transform}) and
(\ref{eq::w-equation}).

\clearpage

\figcaption[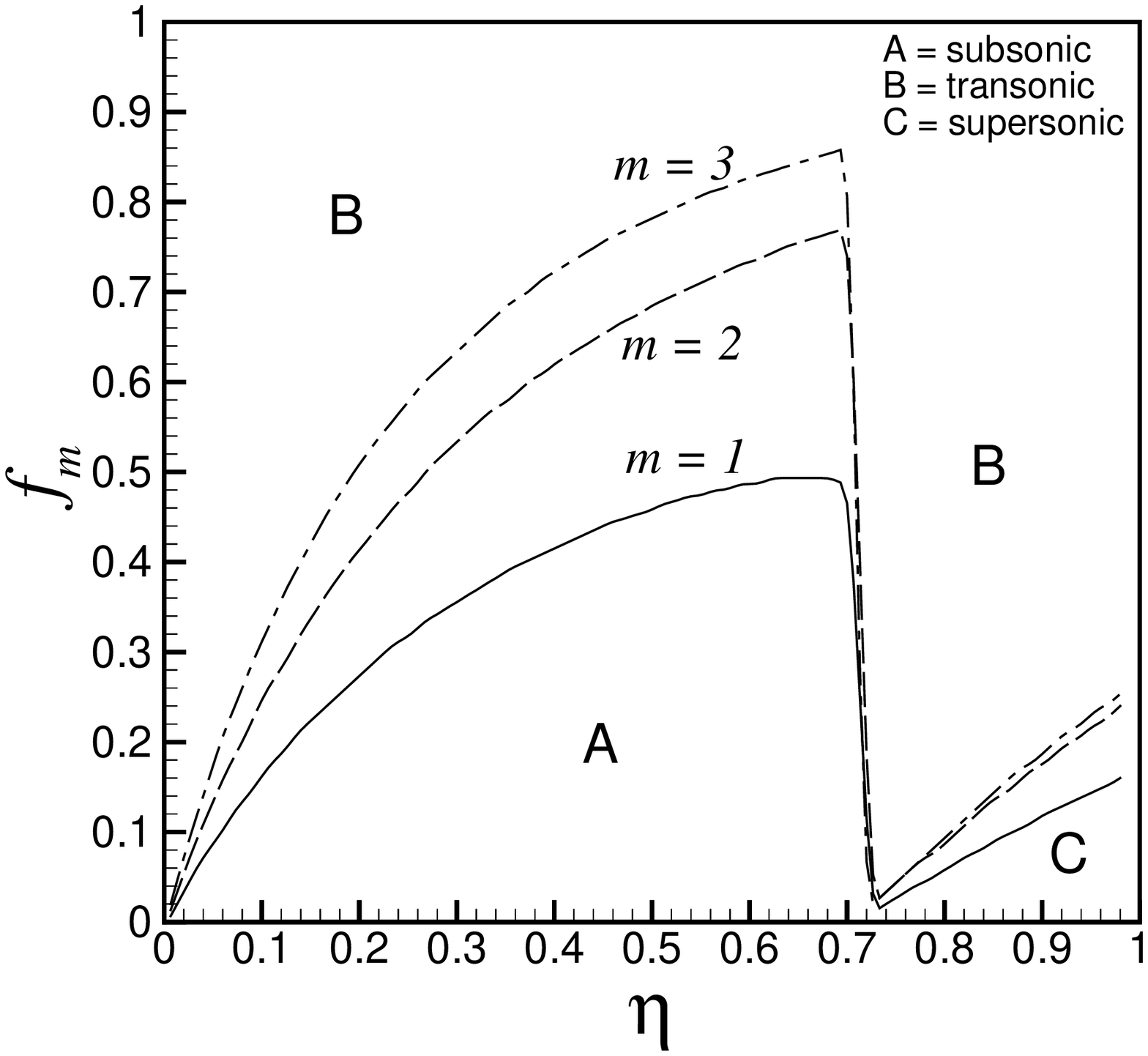]{The $(\eta,f_m)$ parameter space of simple
disks. Solid, dashed, and dash-dotted lines respectively show the
boundary of transonic models for $m=1$, $m=2$, and $m=3$. More
elongated subsonic models are supported as the wavenumber $m$ is
increased. Entirely supersonic models can exist for a limited
ranges of parameters. Such models are almost axisymmetric.
\label{fig1}}

\figcaption[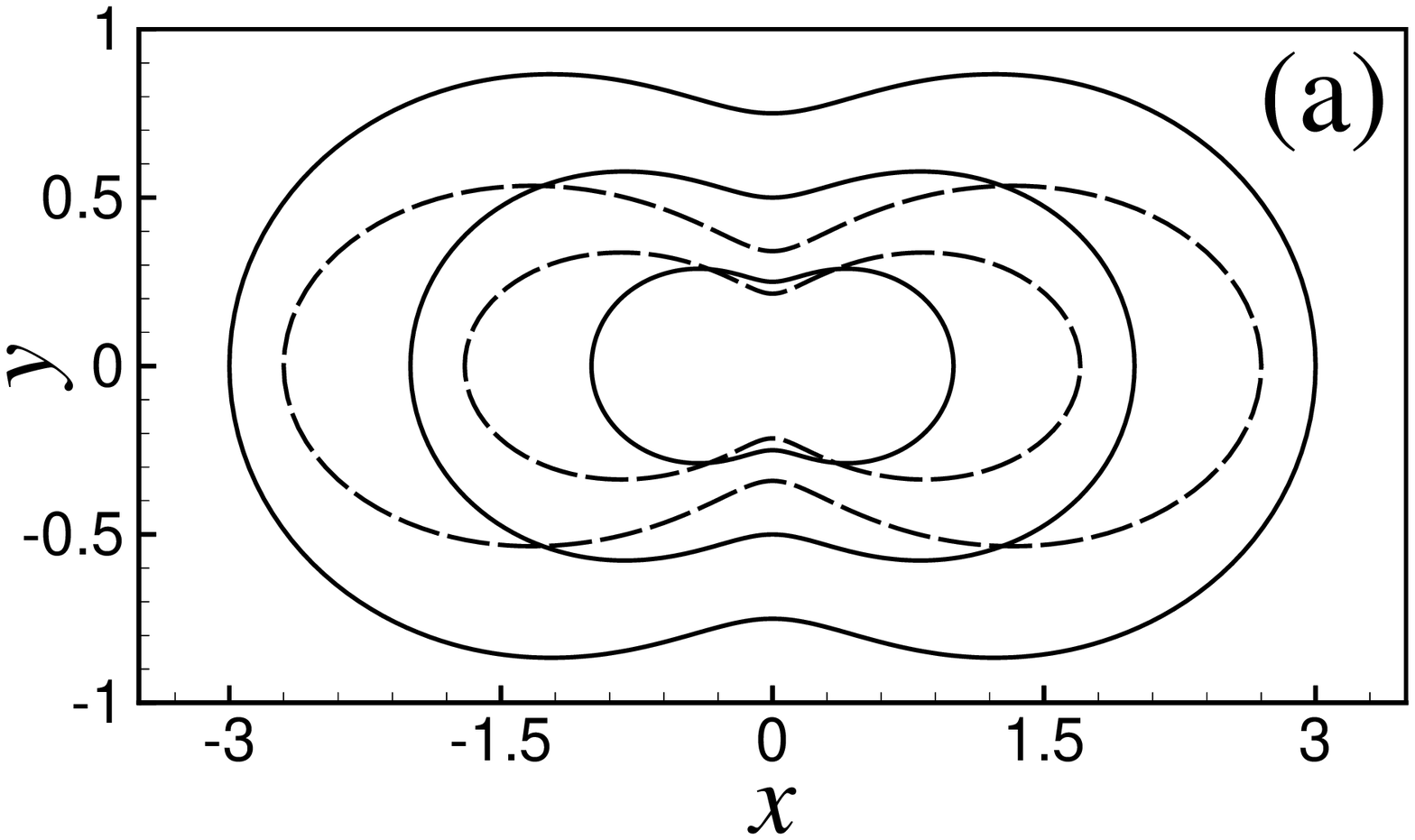,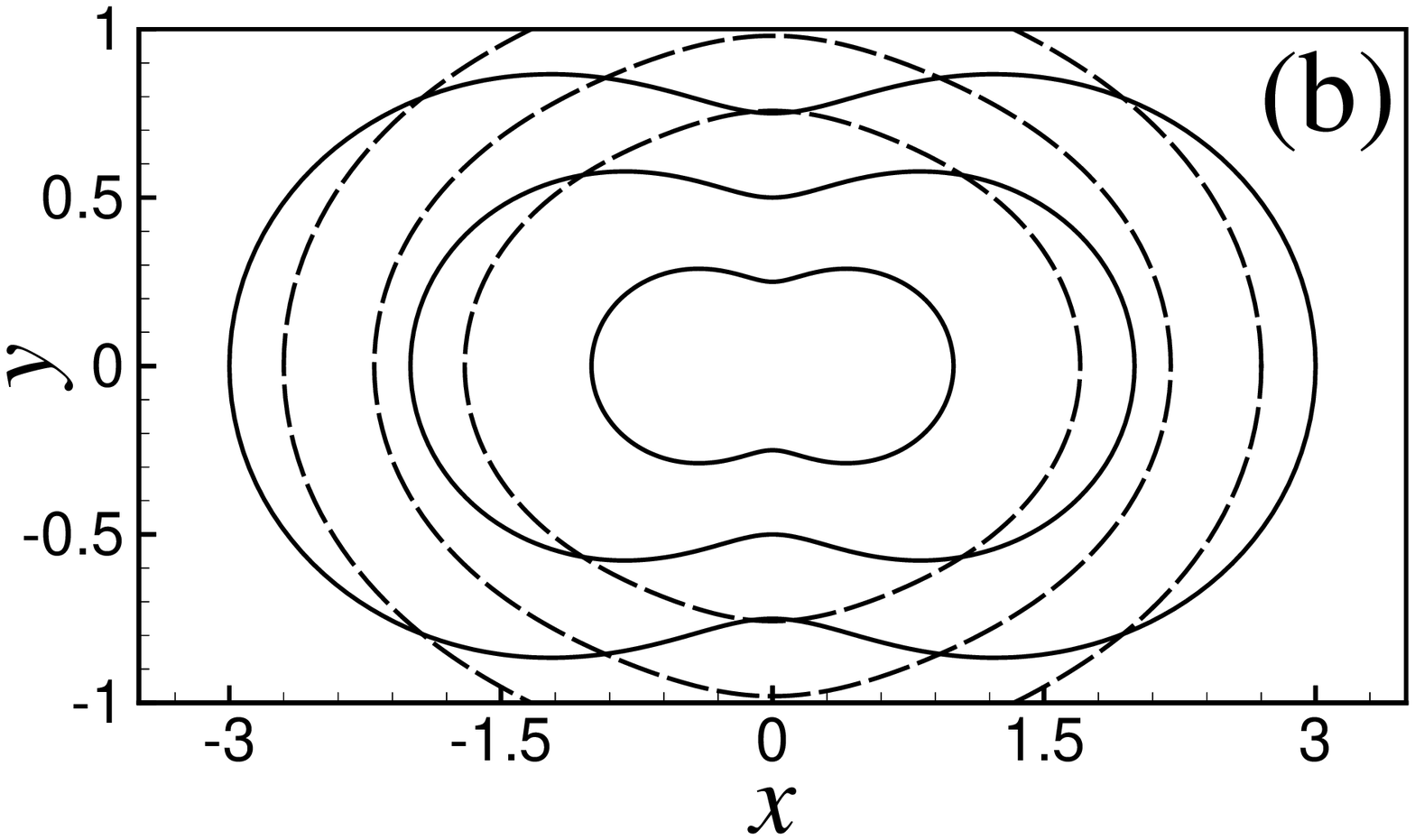,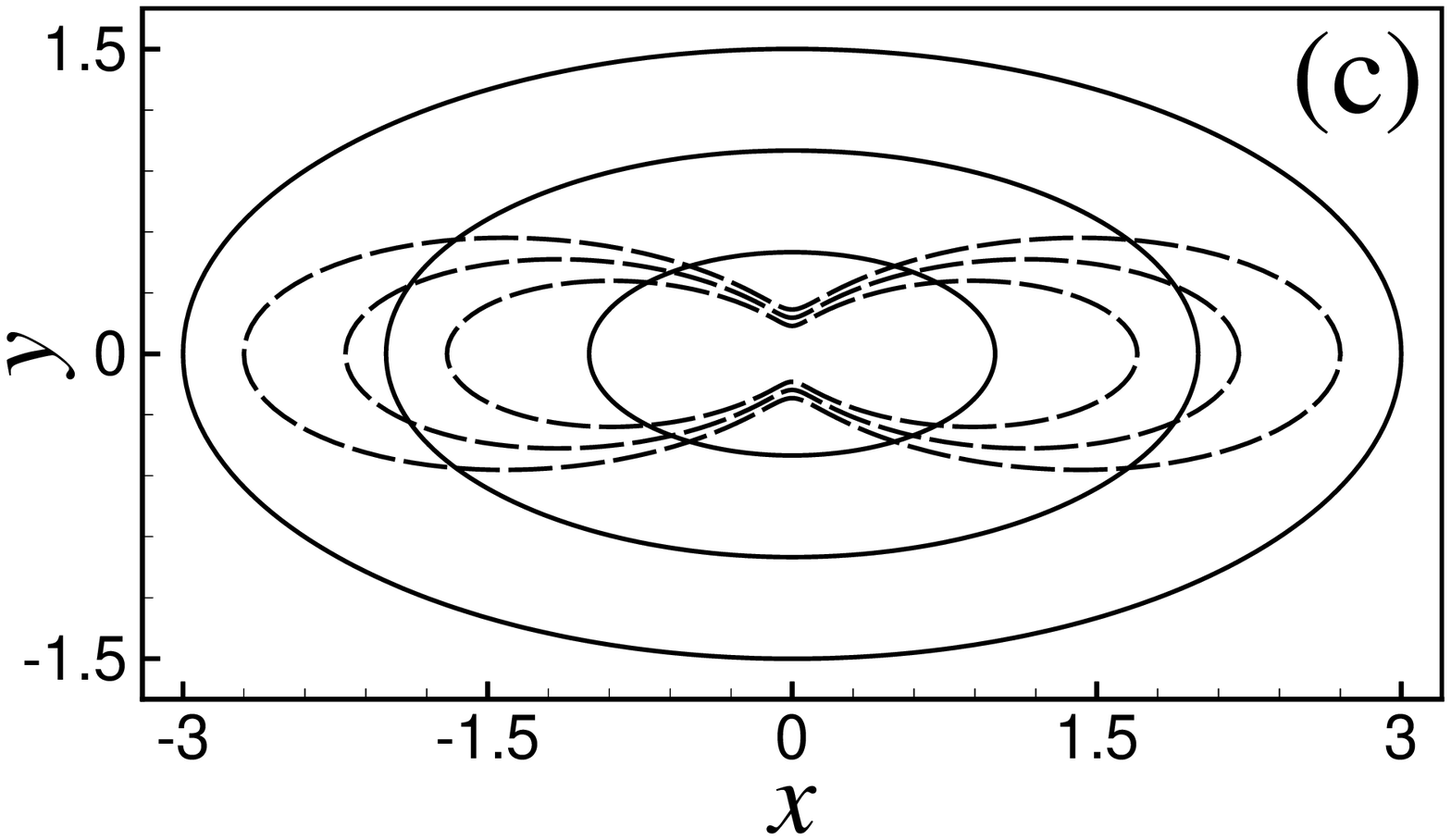,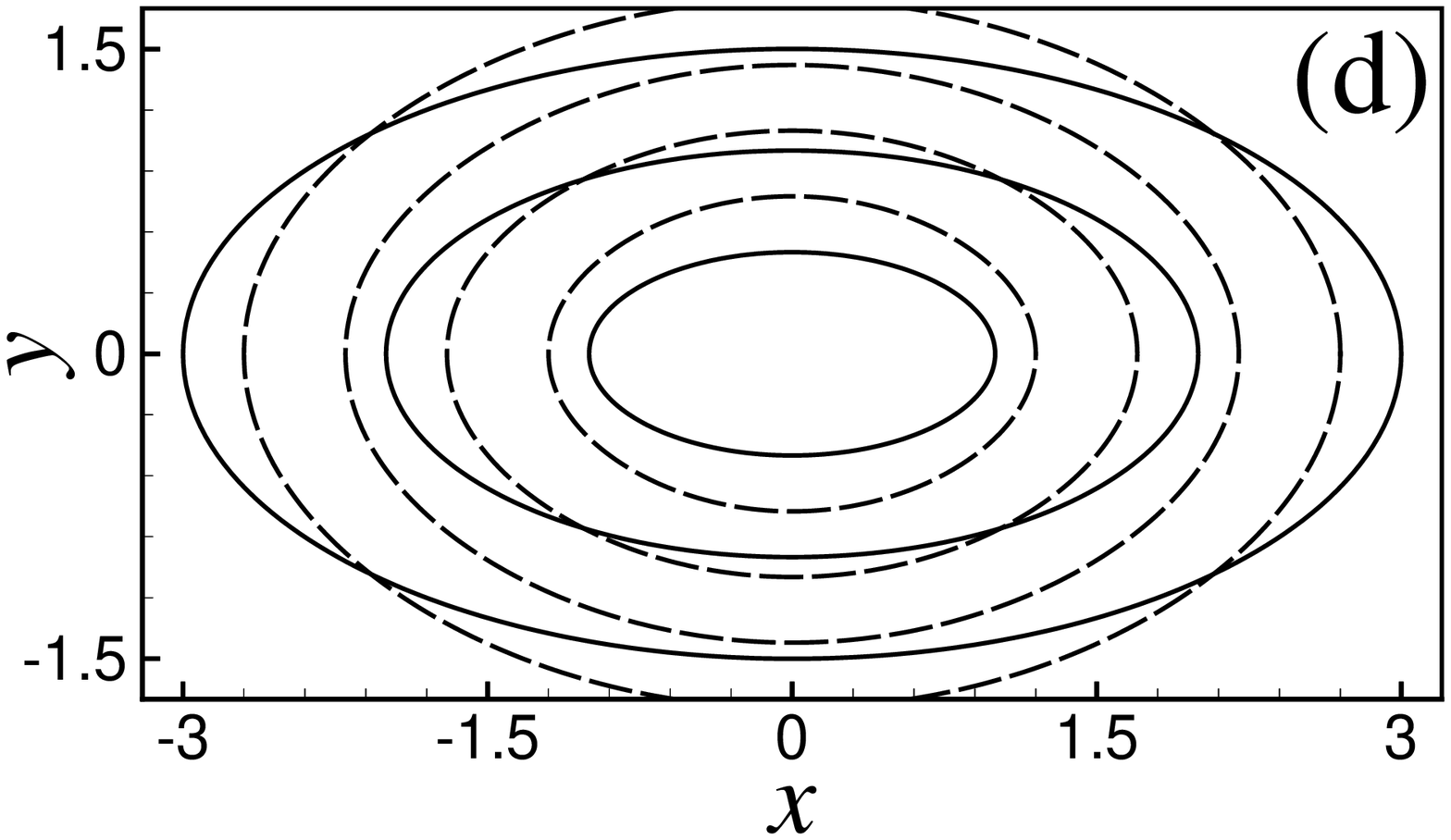]{Isodensity contours
(solid lines) and streamlines (dashed lines) of the power-law and
elliptic disks with flat rotation curves for $\Sigma _0=1$, $\pi
G=1$, and $q=0.5$. ({\em a}) Power-law disk with
$V_{\phi}=\sqrt{A_{\rm min}}/2=0.462$. ({\em b}) Power-law disk
with $V_{\phi}=4\sqrt{A_{\rm min}}/5=0.739$. ({\em c}) Elliptic
disk with  $V_{\phi}=\sqrt{A_{\rm min}}/3=0.581$. ({\em d})
Elliptic disk with $V_{\phi}=4\sqrt{A_{\rm min}}/5=1.395$.
\label{fig2}}

\figcaption[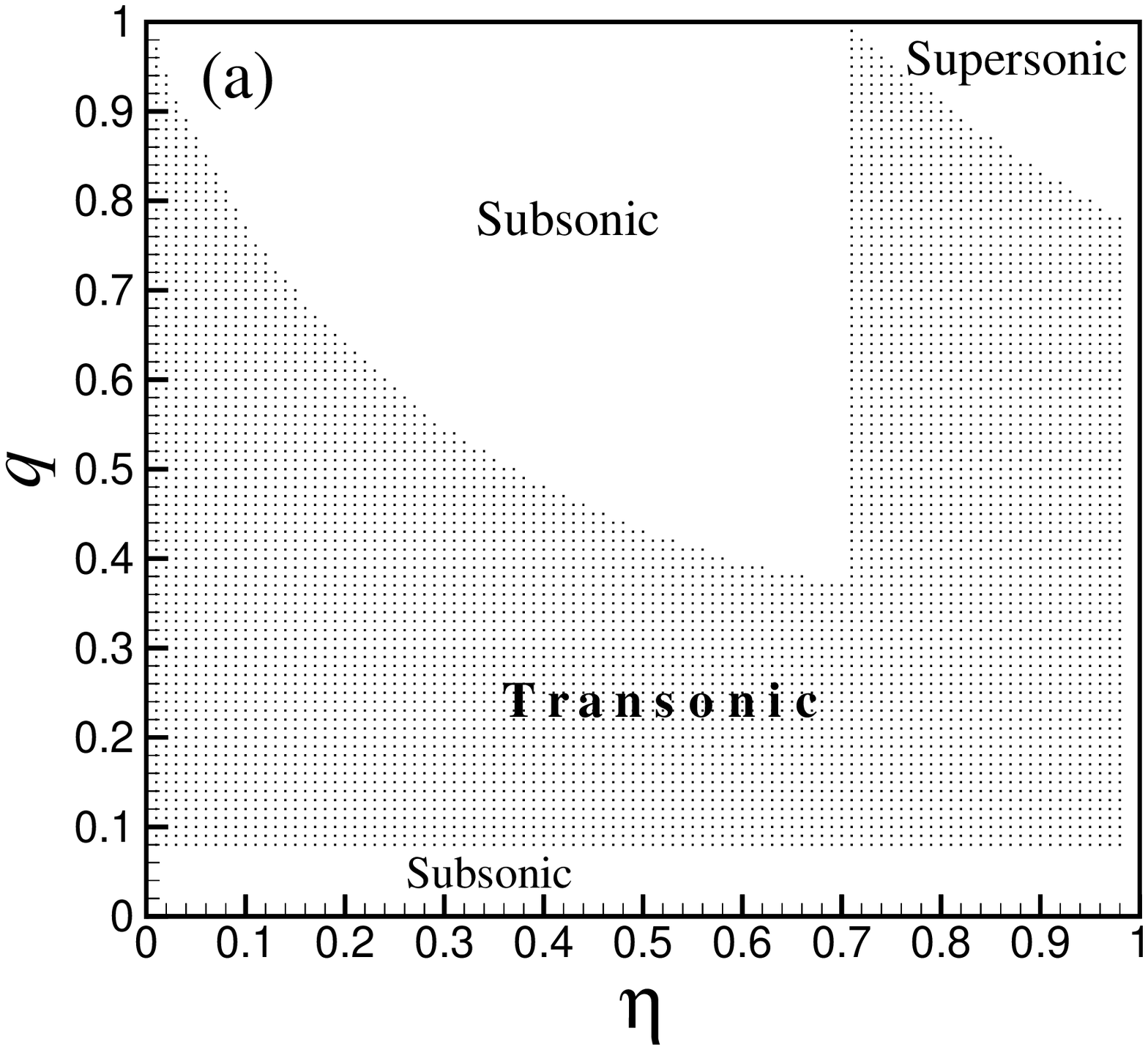,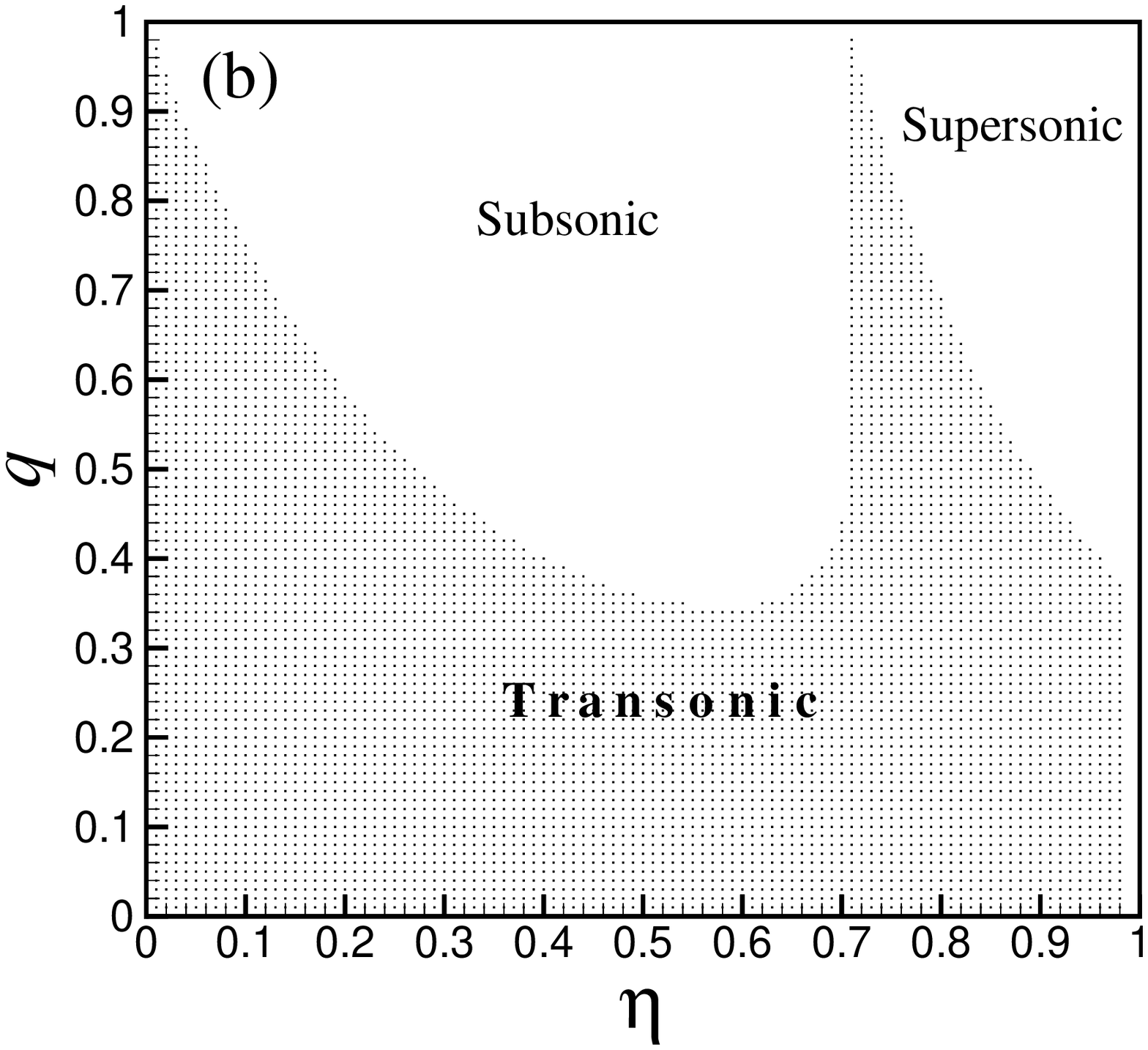]{The $(\eta,q)$ parameter space of the
power-law (Panel {\em a}) and elliptic disks (Panel {\em b}).
Dotted region corresponds to models with transonic flowfields,
which are most likely unphysical. Elliptic disks allow for more
elongated supersonic models than the power-law disks. The maximum
value of $\eta$ taken by subsonic models is equal to $1/\sqrt{2}$
in both disks. \label{fig3}}

\figcaption[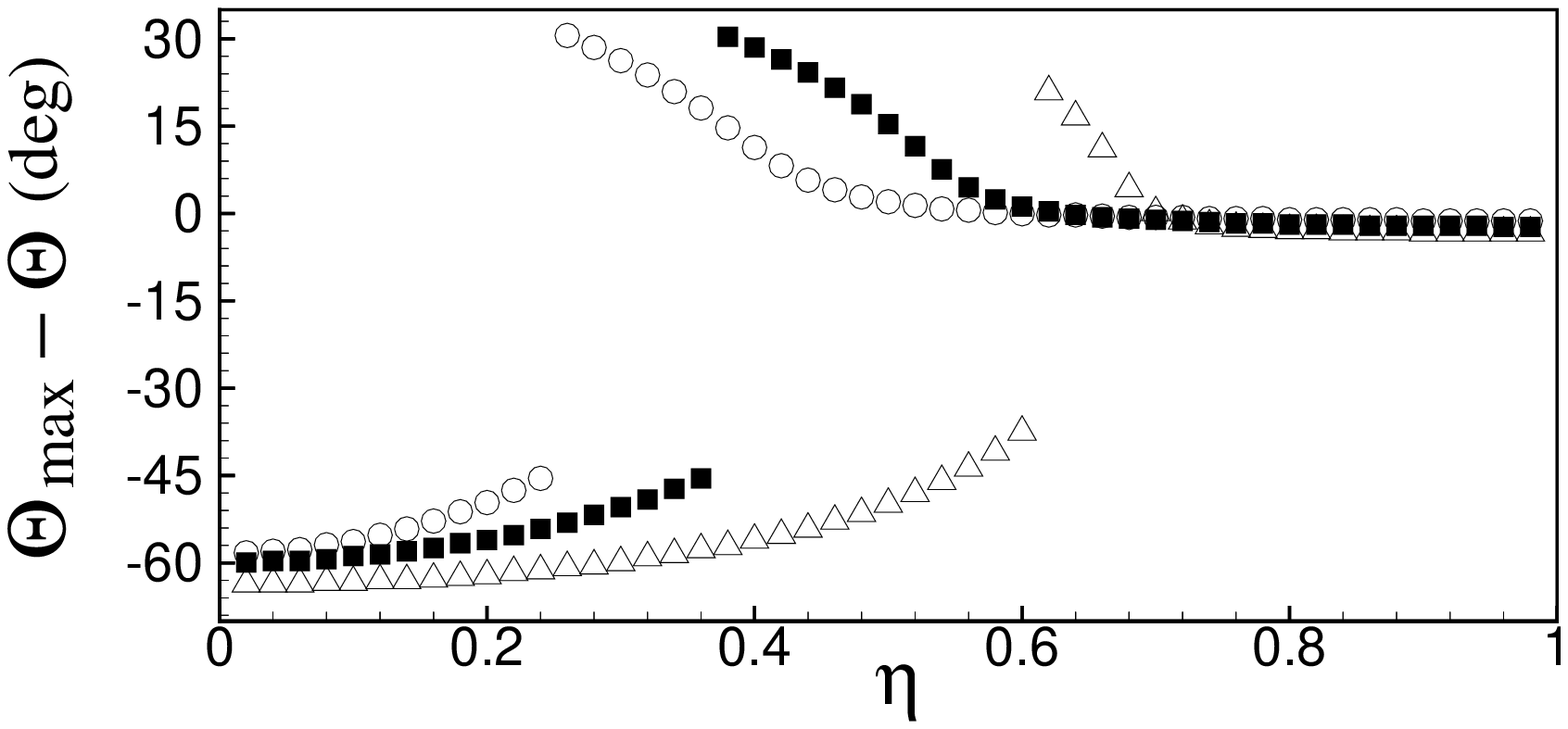]{Variation of $\Delta \Theta =\Theta_{\rm
max}-\Theta$ versus $\eta=1/\sqrt{\beta A_{\rm min}}$ for
elliptic disks, given the viewing angles $\Theta=122^{\circ}.5$
and $i=34^{\circ}.7$. Circles, filled squares, and deltas
correspond to the ellipticities $\epsilon=0.15$, 0.312, and 0.6,
respectively. The existing discontinuities in the data sets are
due to a sudden change in the morphology of streamlines. This
happens at a critical $\eta=\eta_{\rm cr}$. For $\eta<\eta_{\rm
cr}$, streamlines are dimpled.\label{fig4}}

\clearpage

\plotone{f1.ps}

\clearpage

\plotone{f2a.ps}
\plotone{f2b.ps}
\plotone{f2c.ps}
\plotone{f2d.ps}

\clearpage

\plotone{f3a.ps}
\plotone{f3b.ps}

\clearpage

\plotone{f4.ps}

\end{document}